\documentclass[twocolumn]{aastex62}
\usepackage{float}

\graphicspath{{./}{figures/}}

\shorttitle{Photometry of the Uranian Satellites \& the Search for Mab}
\shortauthors{Paradis et al.}

\begin{document}

\title{\textbf{Photometry of the Uranian Satellites with Keck and the Search for Mab}}

\correspondingauthor{Samuel Paradis}
\email{samparadis@berkeley.edu}

\author{Samuel Paradis}
\affil{Department of Electrical Engineering and Computer Sciences, University of California, Berkeley, CA 94720}

\author{Chris Moeckel}
\affil{Department of Earth and Planetary Sciences, University of California, Berkeley, CA 94720}

\author{Joshua Tollefson}
\affil{Department of Earth and Planetary Sciences, University of California, Berkeley, CA 94720}

\author{Imke de Pater}
\affil{Department of Astronomy, University of California, Berkeley, CA 94720}
\affil{Department of Earth and Planetary Sciences, University of California, Berkeley, CA 94720}

{\footnotesize
THE ASTRONOMICAL JOURNAL

\textit{Received 2019 May 14; revised 2019 September 4; accepted 2019 September 5; published 2019 MM DD}
}



\begin{abstract}
 We present photometric properties of six small (radii $<$ 100 km) satellites of Uranus based on 32 \textit{H}-(1.49-1.78 $\mu$m) band images taken on August 29, 2015 from the Keck II Telescope on Maunakea, Hawaii with the near-infrared camera NIRC2 coupled to the adaptive optics system. The sub-observer latitude of our observations was 31$^\circ$, i.e., we view much of the satellites' north poles, in contrast to the 1986 \textit{Voyager} measurements. We derive reflectivities based on mean-stacking measurements of these six minor moons of Uranus. We find that the small satellites are significantly brighter than in previous observations, which we attribute to  albedo variations between hemispheres. We also search for Mab, a small satellite with an unknown surface composition, orbiting between Puck and Miranda. Despite the significantly improved signal-to-noise ratio we achieved, we could not detect Mab. We suggest that Mab is more similar to Miranda, an icy body, than to the inner rocky moons. Assuming Mab is spherical with a radius of 6 km, as derived from \textit{Hubble Space Telescope (HST)} observations if its reflectivity is $\sim$0.46, we derive a 3$\sigma$ upper limit to its reflectivity [\textit{I/F}] of 0.14 at 1.6 $\mu$m. \end{abstract}




\section{Introduction} \label{sec:intro}

Twenty years after NASA's \textit{Voyager} flyby of Uranus, \cite{Showalter} announced the discovery of an outer ring system of Uranus. Deep exposures using the \textit{Hubble Space Telescope (HST)} Advanced Camera for Surveys from July 2003 to August 2005 revealed the existence of two additional dusty rings: the $\nu$-ring and the $\mu$-ring. These rings are located radially outwards of the main ring system. Because both the $\nu$ and $\mu$-rings are brighter in high-phase angle \textit{Voyager} data, \cite{Showalter} suggested that they must be dusty. Keck infrared images successfully detected the $\nu$-ring, but failed to detect the $\mu$-ring. These results helped \cite{dePater2006a} establish that Uranus's outer ring system resembles that of Saturn: a red inner ring (the $\nu$-ring, similar to Saturn's G-ring), and a blue outer ring (the $\mu$-ring, resembling Saturn's E-ring). When the particles that make up the ring are smaller than or comparable to the wavelength of light, the reflected color is dominated by particle size effects rather than the intrinsic color of the material. Dusty rings, therefore, typically have a red color, such as seen on Jupiter (\cite{Neugebauer}; \cite{Nicholson}; \cite{dePater1999}). A blue color is indicative of Rayleigh scattering, i.e., the rings are dominated by grains much smaller than the wavelength of light \citep{dePater2006a}. The HST images also revealed the presence of a new moon within the $\mu$-ring: Mab. Mab orbits in between the orbits of Puck and Miranda \citep{Showalter}. Puck is much darker than Miranda, perhaps suggestive of a more rocky composition, compared to Miranda's icy composition \citep{Karkoschka}. In the present paper, we focus on using Keck infrared observations to constrain Mab's surface composition in order to determine if Mab has a predominately icy or rocky surface. Both a detection and nondetection in the infrared reveal information about the surface, albedo, and size of Mab.

After a discussion of the observations in Section 2.1, we determine the center of Uranus (Section 2.2), used for locating the relevant moons within the system. Next, we stack the exposures to increase the signal-to-noise ratio (Section 3.1), use these results to obtain the reflectivities of the six major moons (Section 3.2), and ultimately constrain Mab's radius and albedo (Section 3.3). Finally, we reflect on the size and albedo constraints of Mab derived from the nondetection in the infrared (Section 4).

\section{Data} \label{sec:data}

\subsection{Observations} \label{subsec:Observations}

We observed Uranus on August 29, 2015 UTC from the Keck II Telescope on Maunakea, Hawaii in the \textit{H}-band (1.49-1.78 $\mu$m). A total of 32 images were taken with the narrow camera of the NIRC2 instrument coupled to the Adaptive Optics system \citep{Wizinowich}. The planet Uranus itself was used for wavefront sensing \citep{vanDam}. The result was a $1024\times1024$ array with a scale of 0.009942 arcsec/pixel \citep{dePater2006b}. The angular resolution was 0.045 arcseconds. Images were taken using an integration time of 120 seconds, which maximizes signal strength while minimizing feature smearing caused by the movement of the moons. All images were processed using standard near-infrared data reduction techniques, i.e., flat fielding, sky subtraction, and replacement of bad pixels with the median of adjacent pixels. Each image was corrected for the geometric distortion of the array using the 'dewarp' routines provided by P. Brian Cameron\footnote{https://www2.keck.hawaii.edu/inst/nirc2/astrometry/nirc2dewarp.pro}, who estimates residual errors at $\leq$ 0.1 pixels. Photometric calibration was preformed on the star HD 1160 (7\textsuperscript{th} mag A-type star; \citeauthor{Elias} 1982): this resulted in a conversion factor of $1.1 \times 10^{-16}$ erg s$^{-1}$ cm$^{-2}$/$\mu$m for 1 count/second.

\subsection{Aligning the Image Data Cube to Find the Center Coordinate} \label{subsec:image_cube}

Ephemerides of the Uranian satellites are obtained from Jet Propulsion Laboratory's  HORIZONS\footnote{https://ssd.jpl.nasa.gov/horizons.cgi}, and are given relative to the center of Uranus, requiring good knowledge of the location of the center for a given image. However, small fluctuations in pointing propagate into uncertainties of the planet's center position, requiring adjustments. We first aligned all observations using a cross-correlation technique. The best estimate for Uranus' center and its uncertainties were then obtained by minimizing the differences between the predicted locations of the moons (as defined by HORIZONS ephemeris data) and the actual locations (as defined by the brightest local pixel) in each image. We did not consider instances in which the detection of the moon was compromised by noise, rings, or other irregularities. Using first-order spline interpolation to obtain subpixel accuracy, we constrained the accuracy of the center pixel to be $\pm .27$ pixels in \textit{x} and $\pm .42$ pixels in \textit{y}. This translates to a total error of .5 pixels. Our technique is described in more detail by \cite{Tollefson} and \cite{Luszcz}. To confirm both our alignment and the location of the center, we stacked all individual frames to produce Fig. 1, and overlaid the moons' expected elliptical orbits on the image. The moons themselves show up as short arcs, traced out during the observations. As shown, these arcs align well by visual inspection, confirming the images are aligned and the center pixel is accurate. 

\begin{widetext}
\begin{center}

\begin{figure}[H]
\centering\includegraphics[width=.90\linewidth]{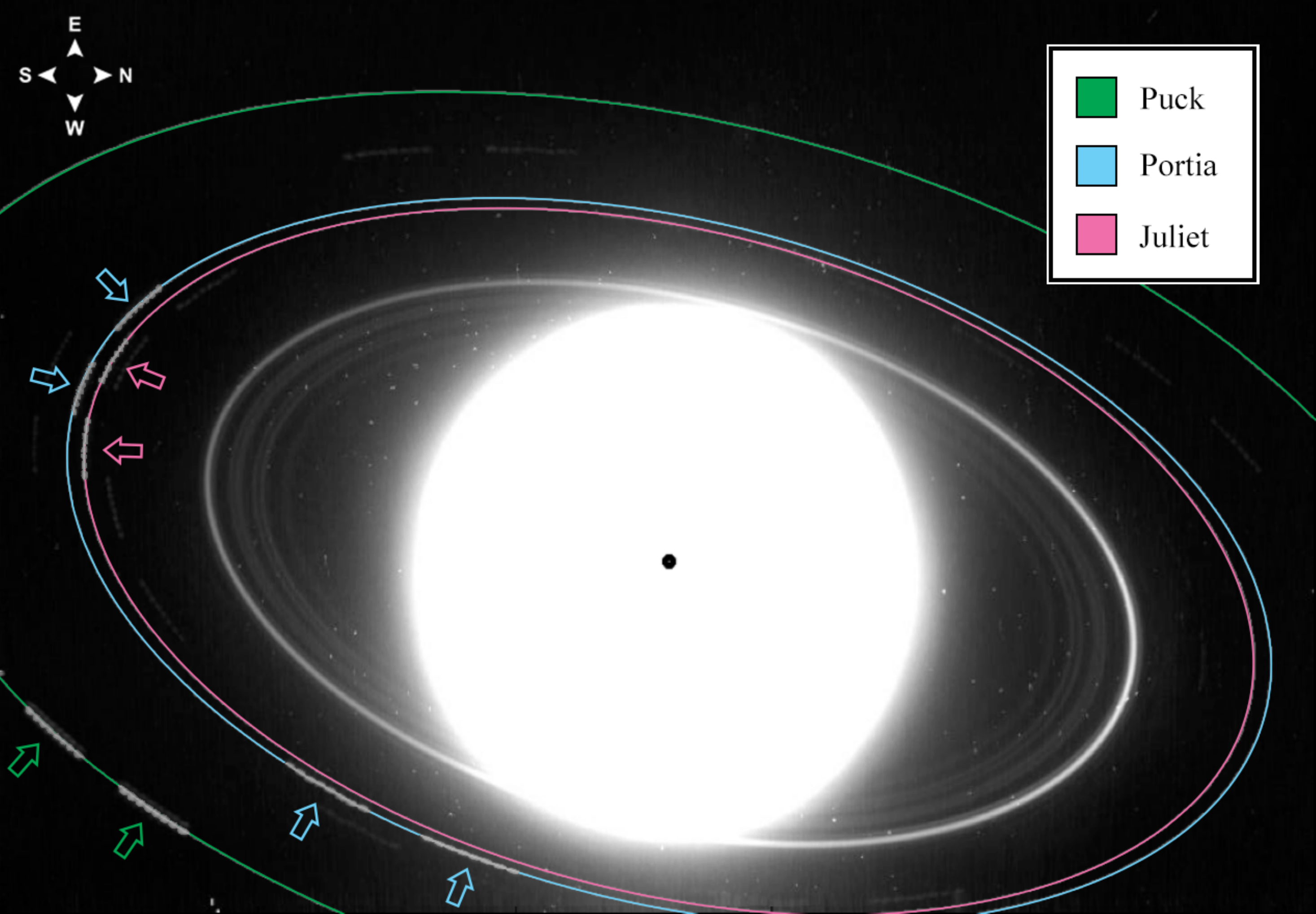}

\caption{Image of Uranus (best estimate center identified by the black dot), its rings, and moons after aligning and co-adding all images. Note that the arcs traced out by the moons are aligned with their expected elliptical orbits. In addition to the moons indicated on the graphic, in between Puck and Portia we see the traces of Rosalind (just below Portia, behind the blue arrows), Belinda (near the top, between the green and blue orbits), and Desdemona (just interior of Juliet, below the lower pink arrow).}
\end{figure}
\end{center}
\end{widetext}

\section{Analysis and Discussion} \label{sec:Analysis}

\subsection{Stacking Moon Images}

At this point, we have a three-dimensional image data cube, with the \textit{x}- and \textit{y}-axes being the right ascension (R.A.) and declination (DEC) of the first image, and the z-axis being time, or the number of images, all  aligned on a common center for Uranus. We create several subimages in the dataset centered on the ephemeris-predicted location of the relevant moons, and mean stack these images to reduce the noise by $\sqrt{N}$, where $N$ is the number of images stacked. Fig. 2 shows an example of the six moons, as detected in one frame, and Fig. 3 shows the result after stacking each moon. Since the moons are essentially point sources (section 3.2), the morphology on the images is essentially that of the point-spread Function (PSF), which varies significantly on timescales of minutes. In some of the images of the brighter moons (e.g., Puck), one can see part of the PSF's Airy ring on the left side of the moon. Movement of the moons across multiple pixels during one exposure results in smearing, which explains the oblong shapes. Slight offsets in the location of a moon from the center of the frame reveal a mismatch with the location as predicted by the ephemeris. Offsets are largest for Desdemona (517 km) and Belinda (592 km). Relatively large offsets were also found for these moons from the HST data, where Belinda's offset is most likely caused by the resonance interaction with Perdita \citep{French}.

\begin{widetext}
\begin{center}
\begin{figure}[H]
\gridline{\fig{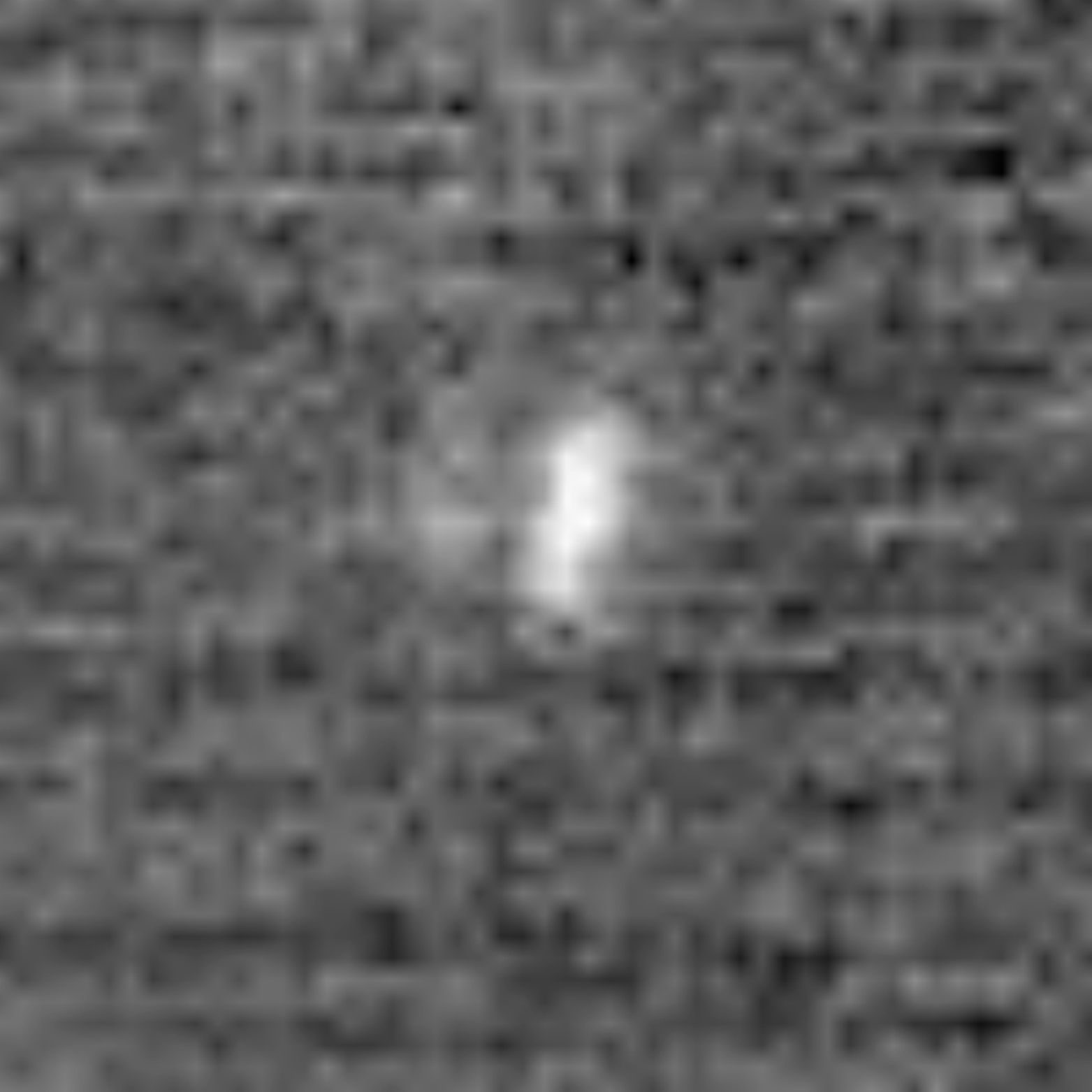}{0.32\linewidth}{Juliet}
          \fig{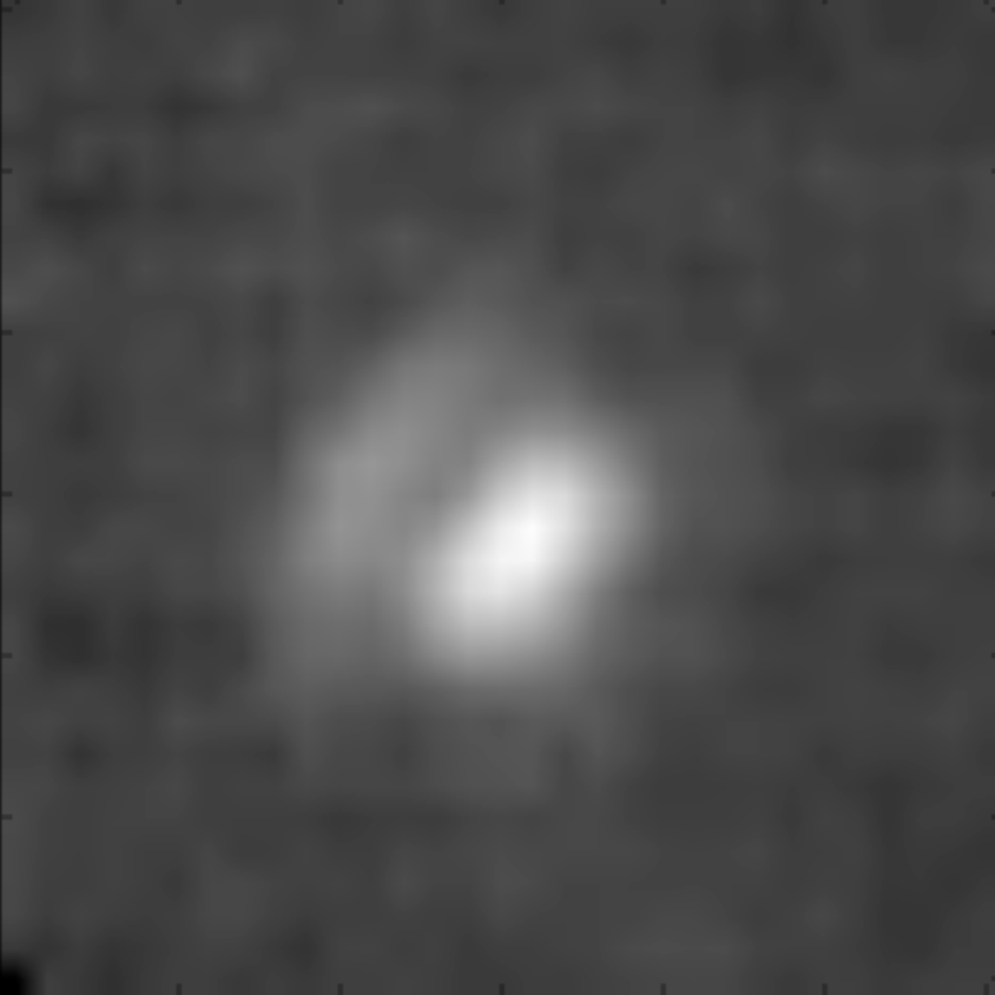}{0.32\linewidth}{Puck}
          \fig{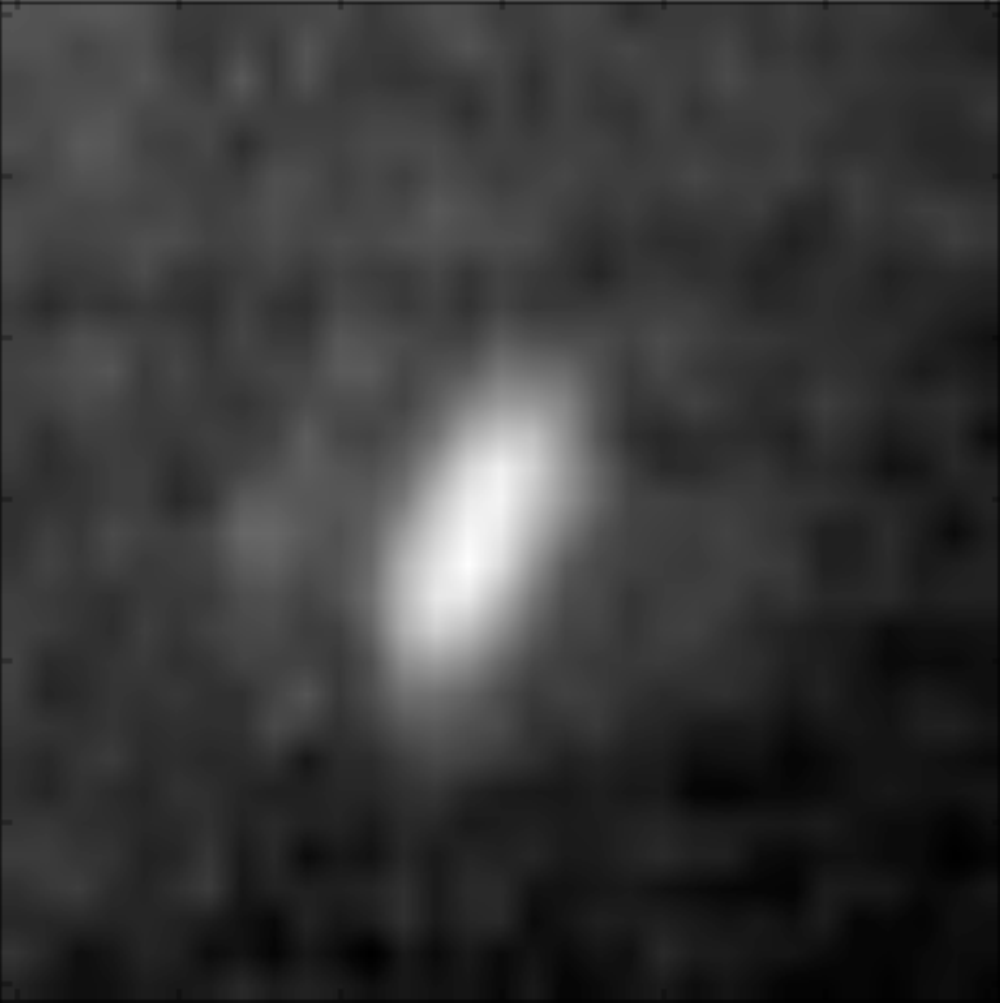}{0.32\linewidth}{Portia}
          }
\gridline{\fig{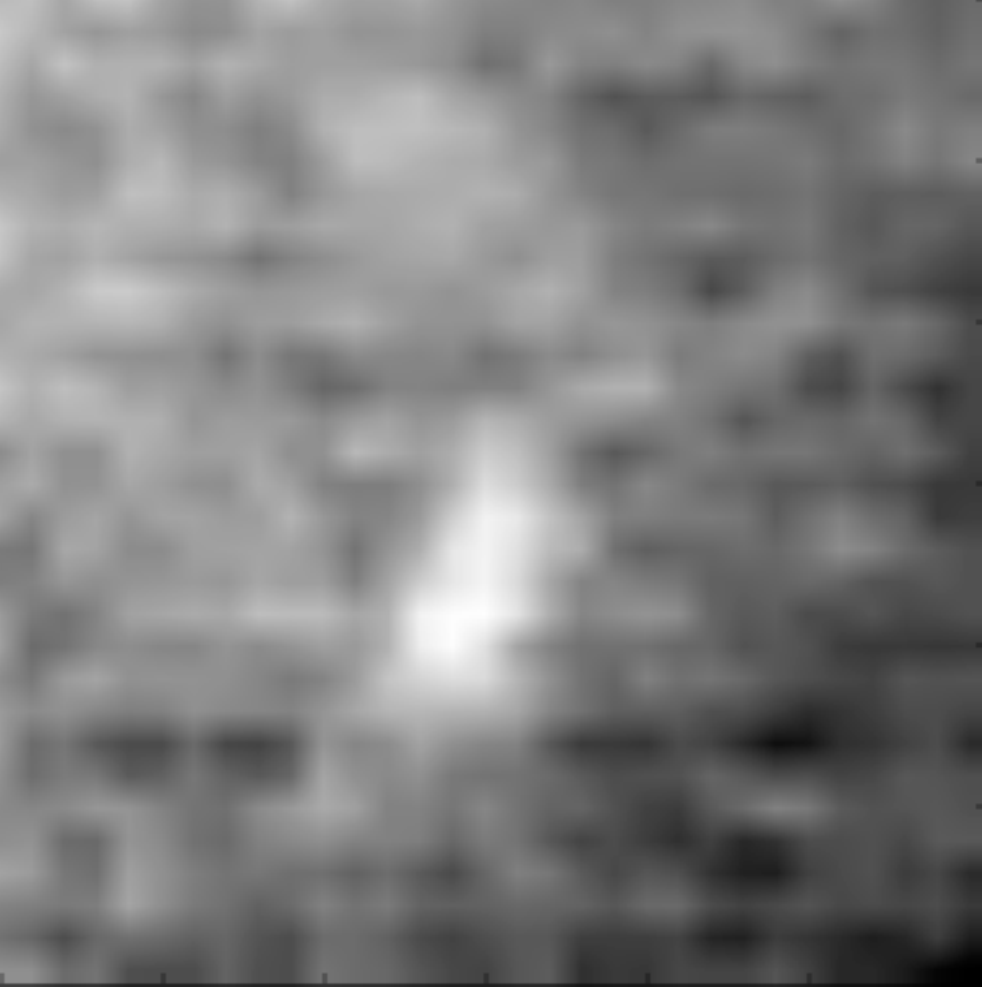}{0.32\linewidth}{Rosalind}
          \fig{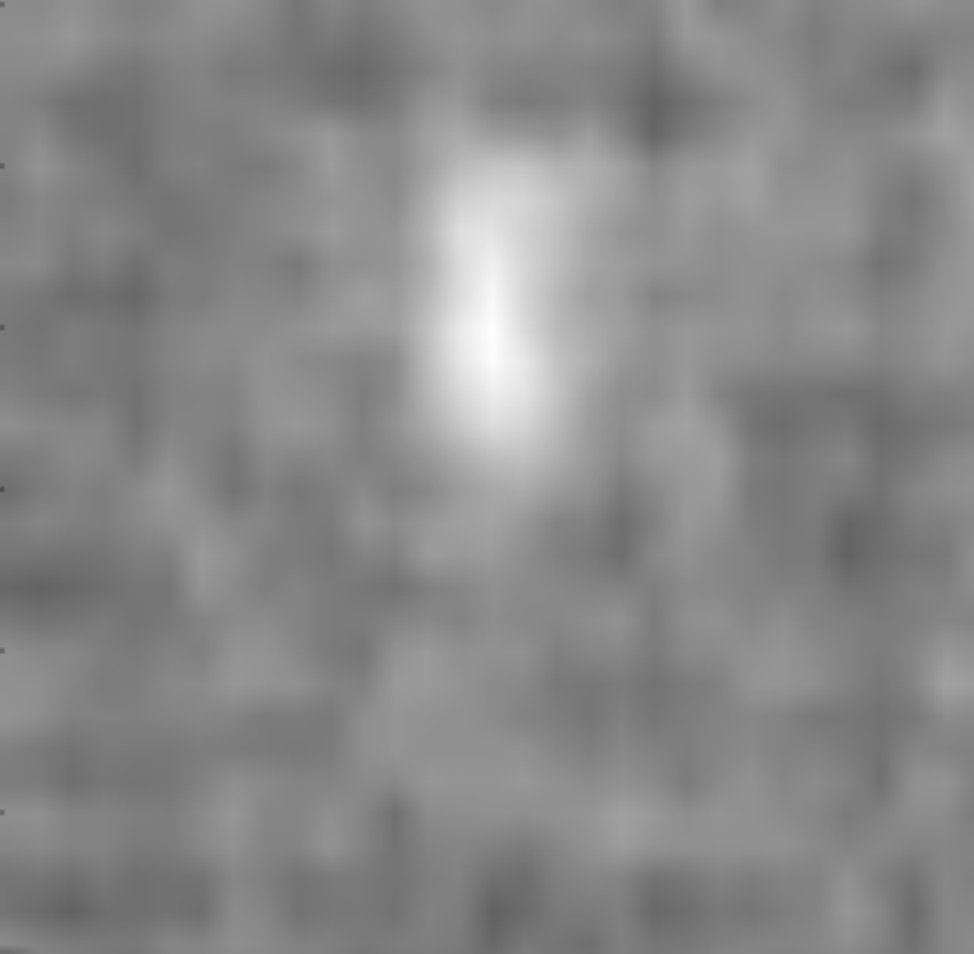}{0.3285\linewidth}{Belinda}
          \fig{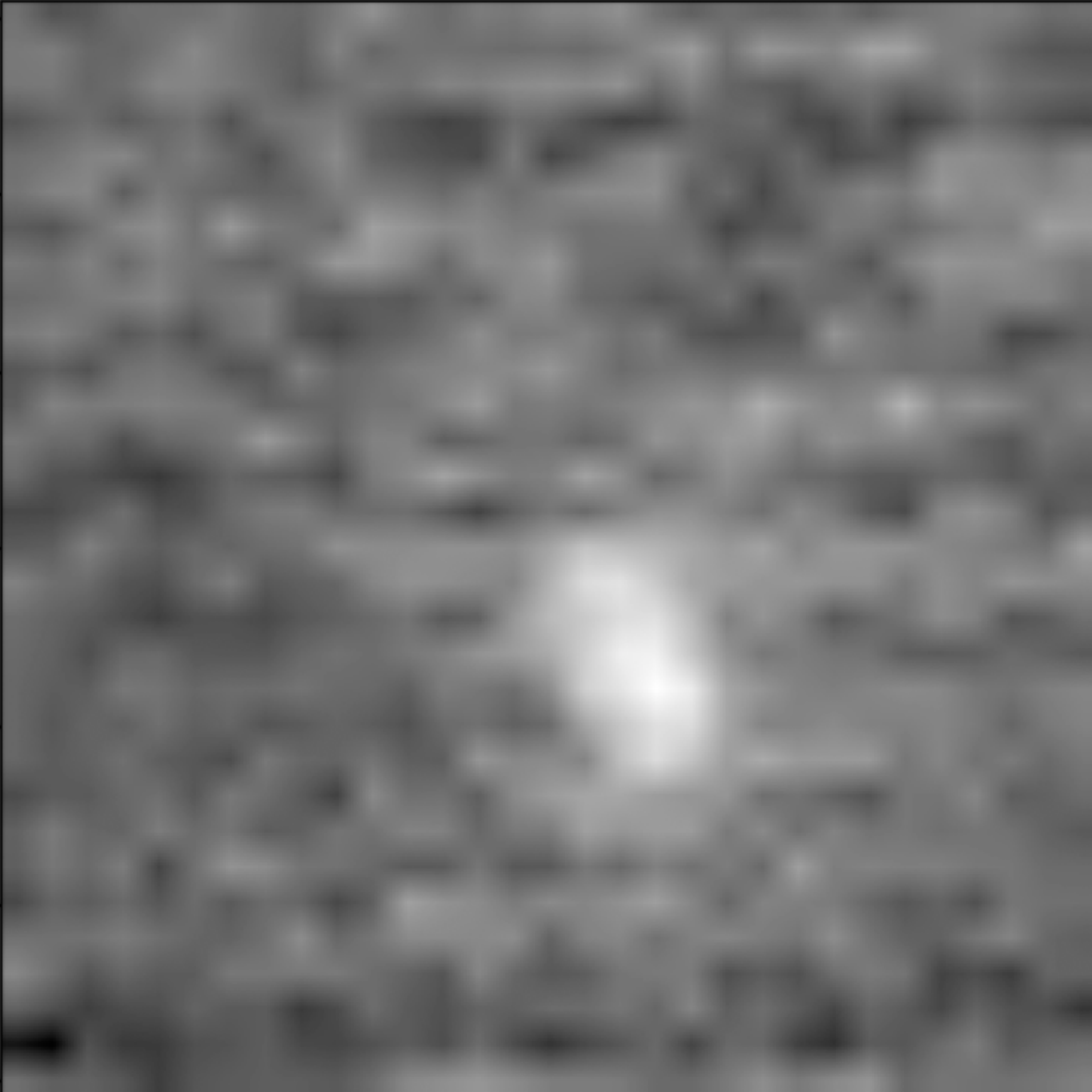}{0.32\linewidth}{Desdemona}
          }
\caption{Example centered images ($30\times30$ pixels) of the six detected moons in our observations before stacking the images; these images are representative of the quality of a single frame. The elongated shape of the moons comes from exposure smearing; the motion of each moon aligns with the longest axis of the residual exposure.}
\end{figure}

\begin{figure}[H]
\gridline{\fig{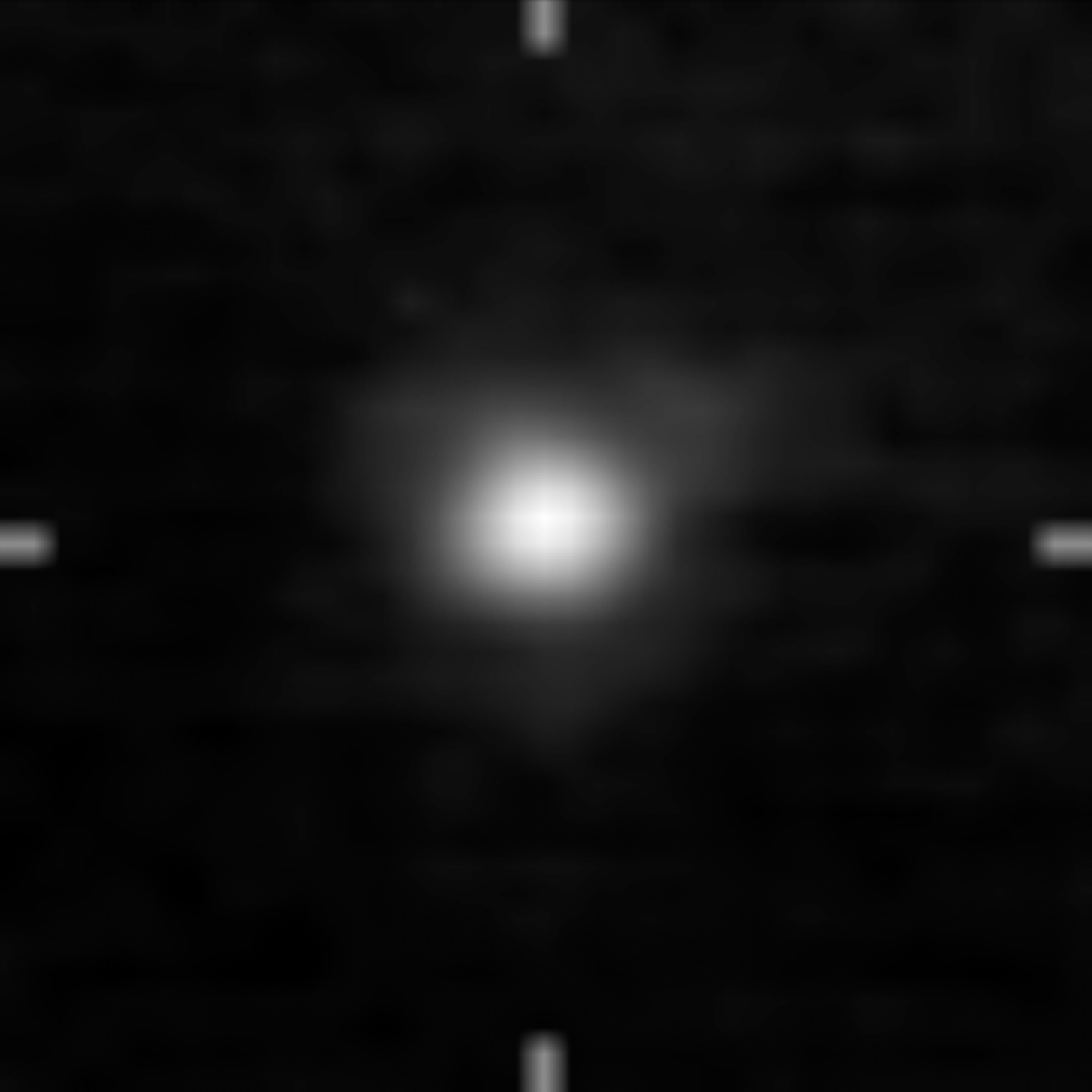}{0.32\linewidth}{Juliet}
          \fig{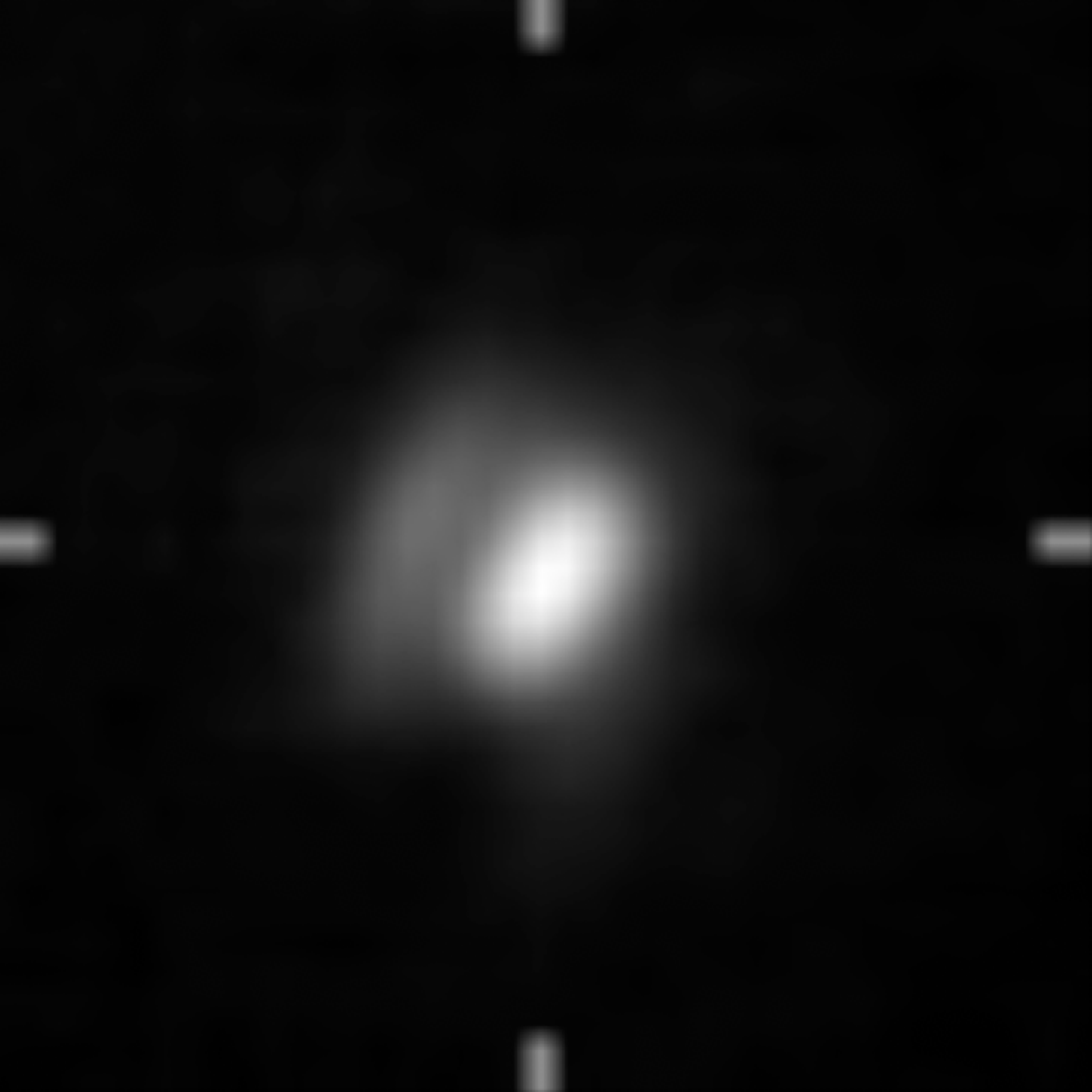}{0.32\linewidth}{Puck}
          \fig{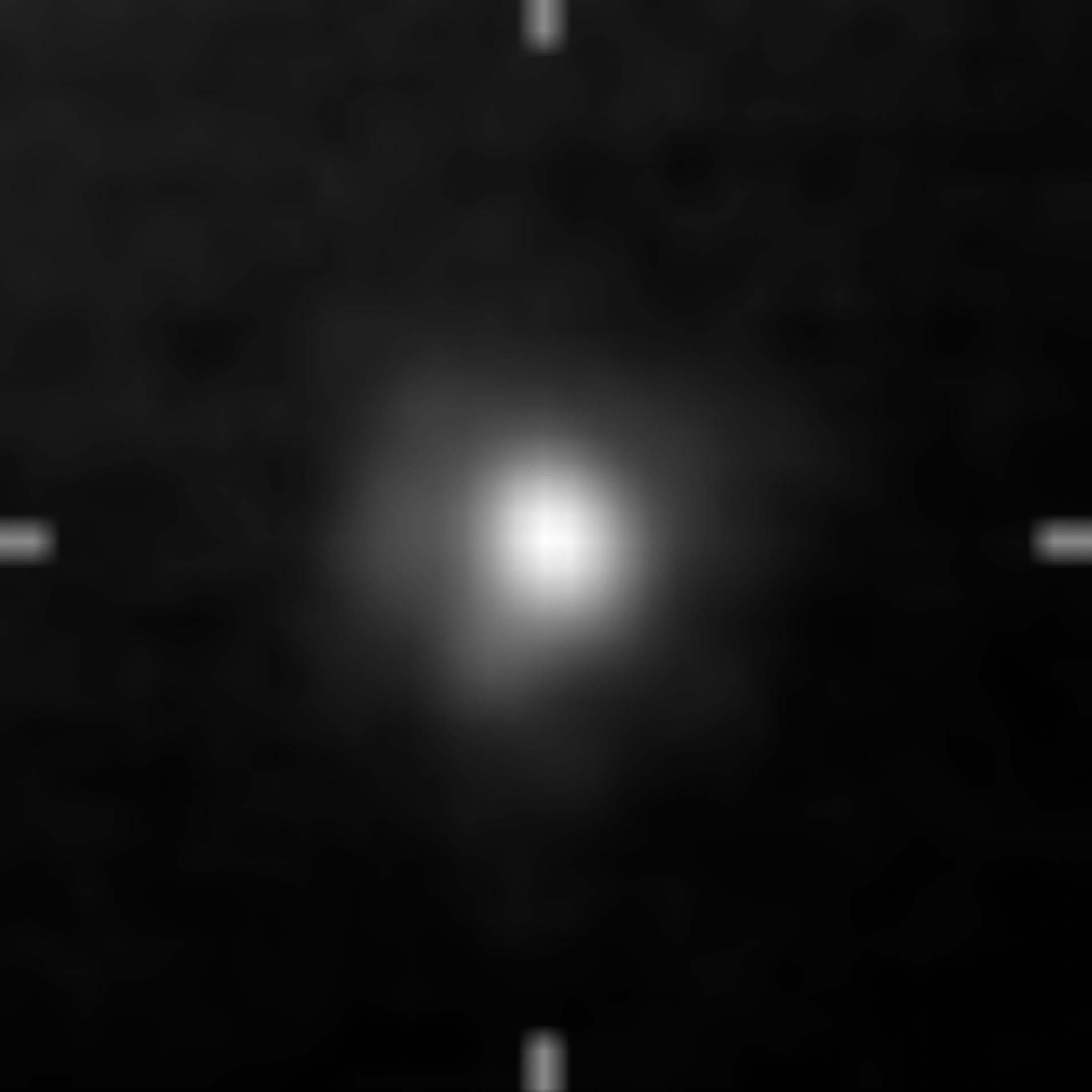}{0.32\linewidth}{Portia}
          }
\gridline{\fig{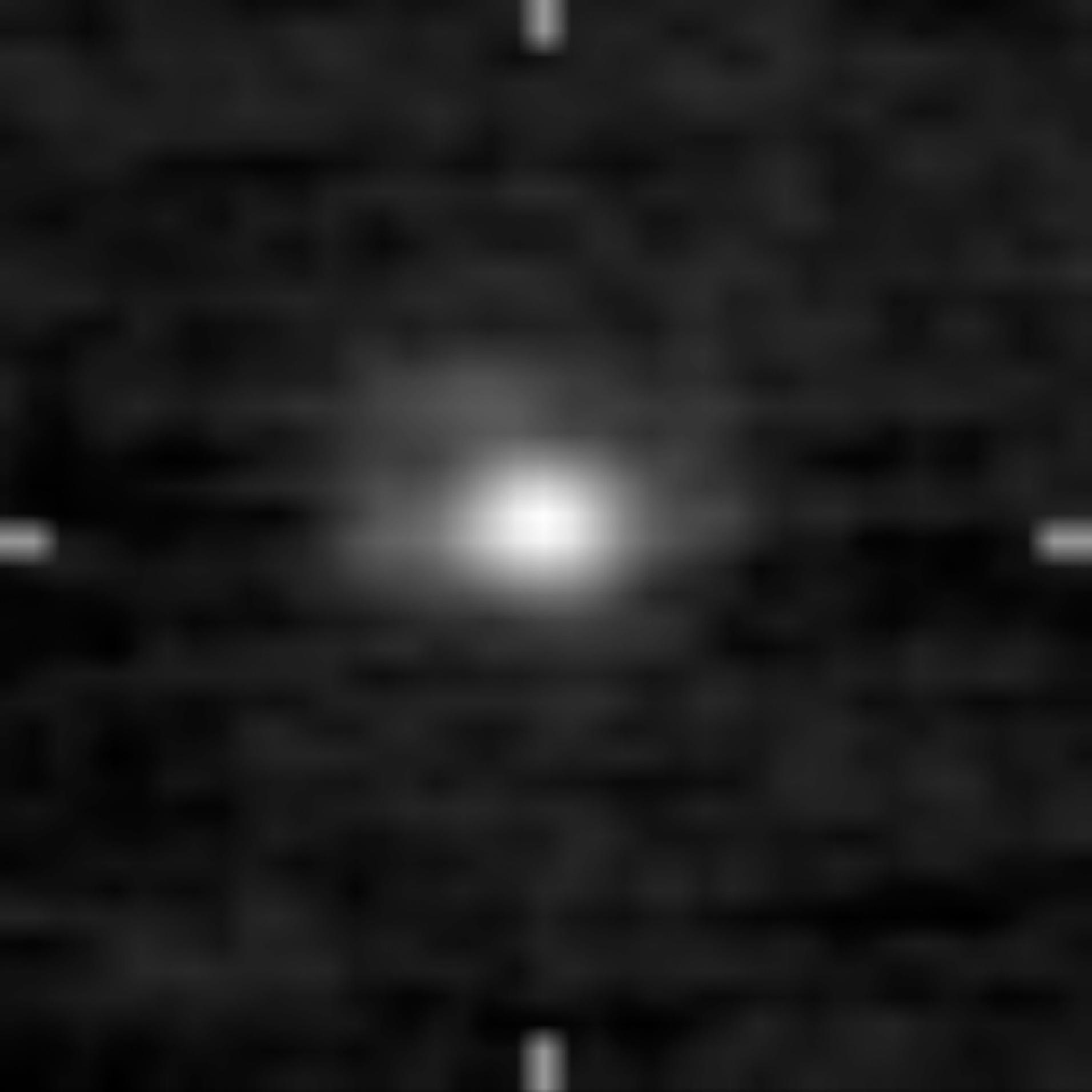}{0.32\linewidth}{Rosalind}
          \fig{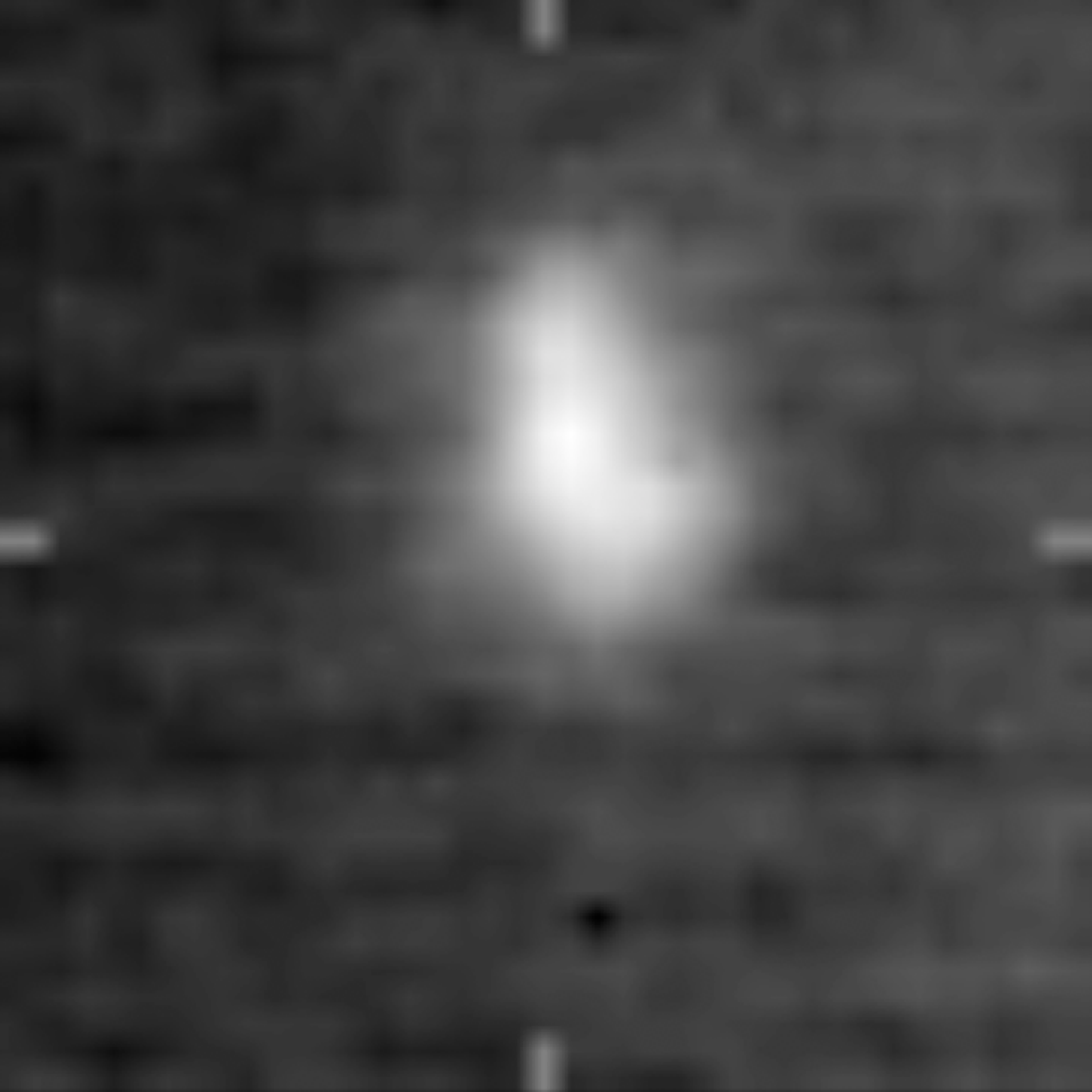}{0.32\linewidth}{Belinda}
          \fig{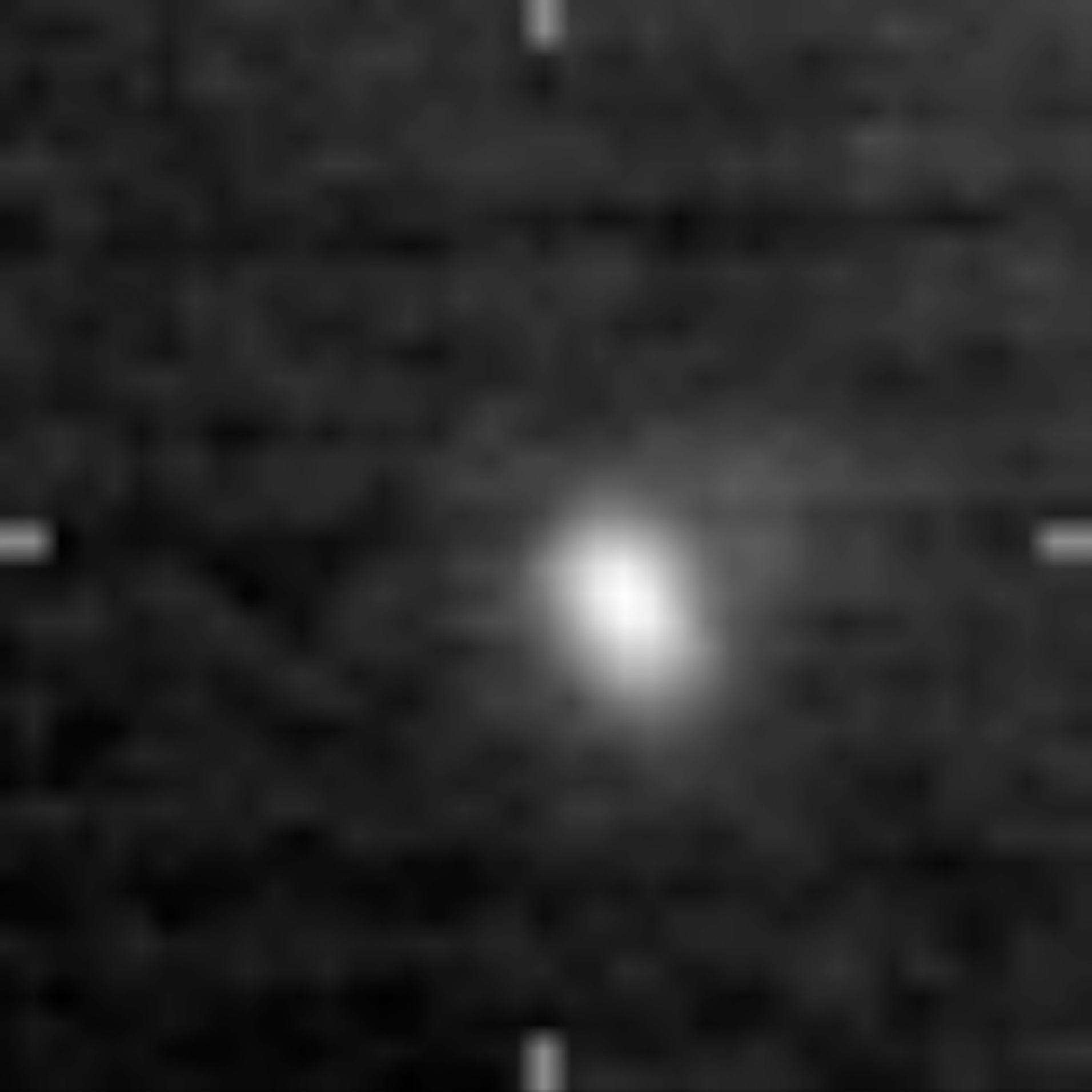}{0.32\linewidth}{Desdemona}
          }
\caption{Post-stacking results of the six moons to illustrate the improvement of the signal-to-noise ratio. The center of the image is the expected position of the moon, marked by the four dashes.  Note that Desdemona is offset from its ephemeris by 3.61 pixels, which translates to 0.0358 arc-seconds (517 km), and the  center of Belinda is offset by 4.12 pixels, which translates to 0.0410 arc-seconds (592 km).}
\end{figure}
\end{center}
\end{widetext}

\subsection{Reflectivities of Puck, Portia, Juliet, Belinda, Desdemona, and Rosalind}

To calculate each moon's albedo, we follow the procedure described by \cite{Gibbard} and \cite{dePater2014} to determine the total amount of reflected light for each moon. We isolated pixels within a specified radius, $r$\textsubscript{inner}, of the maximum value pixel for each image. Next, we performed sky background subtraction by subtracting the median pixel value in an annulus bounded by $r$\textsubscript{inner} and the more distant $r$\textsubscript{outer} from each pixel within $r$\textsubscript{inner}. This process removes the background stray light.

To estimate the total flux, we leverage the fact that both stars and moons are point sources. We first determined the shape of the PSF by using the three observations of our calibrator star. From this we can estimate the amount of flux lost when only observing pixels within $r$\textsubscript{inner}. While a sufficiently large radius ($>50$ pixels) would recover the total flux of the star, a much smaller radius ($<15$ pixels) is needed in order to not integrate over neighboring moons and rings. Considering an  $r$\textsubscript{inner} of 6, 7, 8, 9, 10, 11, and 12 pixels, and an $r$\textsubscript{outer} of 15 pixels, for each $r$\textsubscript{inner}, we subtract the background flux based on $r$\textsubscript{outer}, and then measure the fraction of the total flux of the calibrator star that is captured within $r$\textsubscript{inner}  (see Table 1). For example, when we use an $r$\textsubscript{inner} of 6 pixels, we measure 39.4\% of the total flux from the calibrator star. Thus, we can transform our $F$\textsubscript{detected} for the moons to an adjusted flux, $F$\textsubscript{adjusted}, by dividing $F$\textsubscript{detected} by .394. Calculating this fractional value for the seven radii provides seven different estimations for the flux of a moon; the total flux is the mean of these seven estimations, and the error is the RMS. 

\begin{widetext}
\begin{center}
\begin{deluxetable}{lccccccc}[H]

\tablecaption{Fraction of Total Flux Recovered Using Various Inner Radii ($r\textsubscript{outer} = 15$ Pixels)}
\tablecolumns{8}
\tablenum{1}
\tablehead{
\colhead{$r$\textsubscript{inner} (pixels):} & 
\colhead{6} & 
\colhead{7} & 
\colhead{8} & 
\colhead{9} & 
\colhead{10} & 
\colhead{11} & 
\colhead{12} 
}
\startdata
HD1160-1 & 0.425 & 0.472 & 0.514 & 0.540 & 0.566 & 0.583 & 0.598 \\
HD1160-2 & 0.375 & 0.423 & 0.468 & 0.497 & 0.527 & 0.5455 & 0.561 \\
HD1160-3 & 0.383 & 0.431 & 0.473 & 0.501 & 0.529 & 0.547 & 0.563 \\
\hline
Average & 0.394 & 0.442 & 0.485 & 0.512 & 0.541 & 0.558 & 0.574 \\
\enddata
\tablecomments{These values provide 7 different estimations for the total flux of a moon; the flux is the mean of these 7 estimations, and the error is the RMS. Given a radius of 6 pixels captures $39.4\%$ of total flux, we can calculate the total flux of any point source by calculating the flux in a 6 pixel radius around the point, and dividing by $.394$.}

\end{deluxetable}


\begin{deluxetable}{lccc}[H]
\tablecaption{Transforming Counts/Second to \textit{I/F}}
\tablecolumns{5}
\tablenum{2}
\tablewidth{0pt}
\tablehead{
\colhead{} &
\colhead{$a$ $\times$ $b$} & 
\colhead{$C_1 \times$$F$\textsubscript{adjusted} ($10^{-16}$)} &
\colhead{\textit{I/F} ($10^{-3}$)} \\
\colhead{} &
\colhead{km $\times$ km } & 
\colhead{erg s$^{-1}$ cm$^{-2}$/$\mu$m} &
\colhead{} 
}
\startdata
Puck & 81 $\times$ 81  & 467 $\pm$ 4 & 102 $\pm$ 10\\
Belinda & 64 $\times$ 32  & 96 $\pm$ 2 & 67  $\pm$ 7\\ 
Rosalind & 36 $\times$ 36  & 78 $\pm$ 1 & 86  $\pm$ 9\\
Portia & 78 $\times$ 63  & 290 $\pm$ 2 & 84  $\pm$ 8\\
Juliet & 75 $\times$ 37  & 189 $\pm$ 2 & 97  $\pm$ 10\\
Desdemona & 45 $\times$ 27 & 88 $\pm$ 1 & 104  $\pm$ 10\\
\enddata
\tablecomments{ $a$ and $b$ denote the major and minor radii of the moon  \citep{Karkoschka}.}
\end{deluxetable}

\end{center}
\end{widetext}

For each moon, let $a$, $b$ denote the major and minor semi-axes (i.e., radii), respectively \citep{Karkoschka}. The amount of reflected light, or \textit{I/F}, is calculated using the following formula:

$$\frac{I}{F} = \frac{F\textsubscript{adjusted} C_1}{F_\odot\Omega}$$

\quad where $F\textsubscript{adjusted}$ is the total flux of the moon in counts per second, which is converted to erg s$^{-1}$ cm$^{-2}$/$\mu$m by multiplying by $C_1$, which is obtained from our photometric measurements of the star HD 1160. The factor $\pi  F_\odot$ represents the solar flux at Uranus, and $\Omega$ is the solid angle of moon at geocentric distance $d$: $\frac{\pi {ab}}{d^2}$. Table 2 provides the \textit{I/F} for each moon.

\newpage

Table 3 compares all available infrared observations to date for the moons. While our values match those of \cite{Karkoschka}, we see large discrepancies with \cite{Gibbard}. Since the wavelengths of the observations are essentially the same, differences in reflectivity can be caused by differences in phase angle or geometry. Many of the moons are oblate spheroids, and thus a change in geometry may cause a change in the apparent size. The apparent latitude plays a role in the projected area of the moon, with higher sub-observer latitudes projecting a larger silhouette and vice versa. Assuming the long axis of the moon points toward Uranus due to tidal forces, we can calculate the projected area of each moon. For the \cite{Gibbard} observations, we find a decrease in projected area by a factor of 1.05 compared to ours; the \cite{Karkoschka} observations had a negligible difference in projected area compared to our data. This clearly does not explain the large differences between our data and \cite{Gibbard}. Another factor that is contributing to the \textit{I/F} values is the phase angle. The \cite{Karkoschka} findings outline the \textit{I/F} values for $\lambda = .9 \mu$m for .08$^\circ$ and 1.8$^\circ$. By assuming the impact of differing phase angles to scale uniformly across all wavelengths, we can estimate the effect of phase angle on the \textit{I/F}. For example, given the \cite{Karkoschka} \textit{I/F} value for Rosalind of 0.134 at 1.8$^\circ$ and 0.206 at 0.08$^\circ$, we estimate that $1.56 \times$\textit{I/F}\textsubscript{2.02$^\circ$} $\approx$  \textit{I/F}\textsubscript{0.08$^\circ$}. Using the \cite{Karkoschka} phase angle dependence for each moon and adjusting for the viewing geometry as described above, we change the \cite{Karkoschka} and \cite{Gibbard} measurements to the phase angles and viewing geometries as observed by us; these results are shown in Table 4.

\begin{widetext}
\begin{center}
\begin{deluxetable}{lccccccccc}[H]
\tablecaption{Uranus Moon Reflectivities in the Near-infrared ($10^{-3}$)}
\tablecolumns{10}
\tablenum{3}
\tablewidth{\linewidth}
\tablehead{
\colhead{} &
\colhead{Date} &
\colhead{Phase} &
\colhead{$\mu$m-range} &
\colhead{Juliet} &
\colhead{Portia} &
\colhead{Puck} &
\colhead{Rosalind} &
\colhead{Belinda} &
\colhead{Desdemona}
}
\startdata
This Paper & 8/29/15 & 2.02 - 2.03  &  1.49 - 1.78  & 97 $\pm$ 10 & 84 $\pm$ 8 & 102 $\pm$ 10 & 86 $\pm$ 9 & 67 $\pm$ 7 & 104 $\pm$ 10 \\
\cite{Gibbard} & 10/5/03 & 1.84 - 1.96 & 1.49 - 1.78 & 47 $\pm$ 21 & 42 $\pm$ 7 & 74 $\pm$ 14 & 55 $\pm$ 17 & 43 $\pm$ 20 & -  \\
\cite{Karkoschka} & 7/28/97 & 0.081 & 1.39 - 1.70 & 87 $\pm$ 8 & 80 $\pm$ 9 & 101 $\pm$ 5 & 91 $\pm$ 17 & 77 $\pm$ 11  & 86 $\pm$ 35 \\   
\enddata
\end{deluxetable}

\begin{deluxetable}{lccccccccc}[H]
\tablecaption{Uranus Moon Reflectivities in the Near-infrared ($10^{-3}$) after Adjusting for Different Viewing Geometries and Phase Angles}
\tablecolumns{9}
\tablenum{4}
\tablewidth{0pt}
\tablehead{
\colhead{} &
\colhead{Date} &
\colhead{$\mu$m} &
\colhead{Juliet} &
\colhead{Portia} &
\colhead{Puck} &
\colhead{Rosalind} &
\colhead{Belinda} &
\colhead{Desdemona}
}
\startdata
This Paper & 8/29/15 & 1.63  & 97 $\pm$ 10 & 84 $\pm$ 8 & 102 $\pm$ 10 & 86 $\pm$ 9 & 67 $\pm$ 7 & 104 $\pm$ 10 \\
\cite{Gibbard} & 10/5/03 & 1.63 & 50 $\pm$ 22 & 42 $\pm$ 7 & 74 $\pm$ 14 & 55 $\pm$ 17 & 46 $\pm$ 21 & -  \\
\cite{Karkoschka} & 7/28/97 & 1.59 & 66 $\pm$ 6 & 70 $\pm$ 8 & 80 $\pm$ 4 & 59 $\pm$ 11 & 50 $\pm$ 7 & 55 $\pm$ 22 \\ 
\enddata
\end{deluxetable}
\end{center}
\end{widetext}

A comparison of the values in Table 4 shows that \cite{Gibbard} and \cite{Karkoschka} have quite similar values for Juliet, Puck, Rosalind, and Belinda, while ours are at least 20\% larger for allsixmoons. Only Portia is consistent within the uncertainties of a previous observation \citep{Karkoschka}. One possible explanation and the conclusion that we favor is albedo variations between hemispheres. While the sub-observer latitude for our data was 31$^\circ$, Gibbard's was -18$^\circ$, and Karkoschka's was -40$^\circ$. Thus, it is possible the moons have hemispherical albedo variations. If the northern hemisphere is more reflective than the southern hemisphere, we would expect our \textit{I/F} to be higher than that of \cite{Gibbard} and \cite{Karkoschka}, which is what we see for nearly all of our targets. We acknowledge that it may be hard to imagine why the north poles of all satellites would be brighter; thus we considered the effect of the magnetospheric environment on the moons.
 
Hemispheric differences in albedo on satellites such as Iapetus have been explained by preferential intercepting particles on the leading or trailing hemisphere \citep{Squyres}, depending on whether the moon orbits the planet faster or slower than the planet's magnetic field, which is tied to the rotation of the planet. All moons in our study except for Puck orbit the planet faster than the planet rotates. Yet, we find all moons are brighter, including Puck. Thus, it is difficult to explain a similar method to create a hemispheric albedo difference for all moons this way. Perhaps a hemispheric difference between the two poles may have been established over time through a complex interaction between the solar wind and the Uranian magnetosphere while Uranus orbits the Sun.

\subsection{Search for Mab}

We applied the same methodology of mean-stacking images to Mab. Unfortunately, despite the improved signal-to-noise ratio of the mean-stacked images, we cannot detect Mab. While Fig. 3 shows that other moons are clearly visible after mean-stacking the aligned images, Mab is not discernible in a mean-stacked image, as shown in Fig. 4.

\begin{figure}[H]
\centering\includegraphics[width=.9\linewidth]{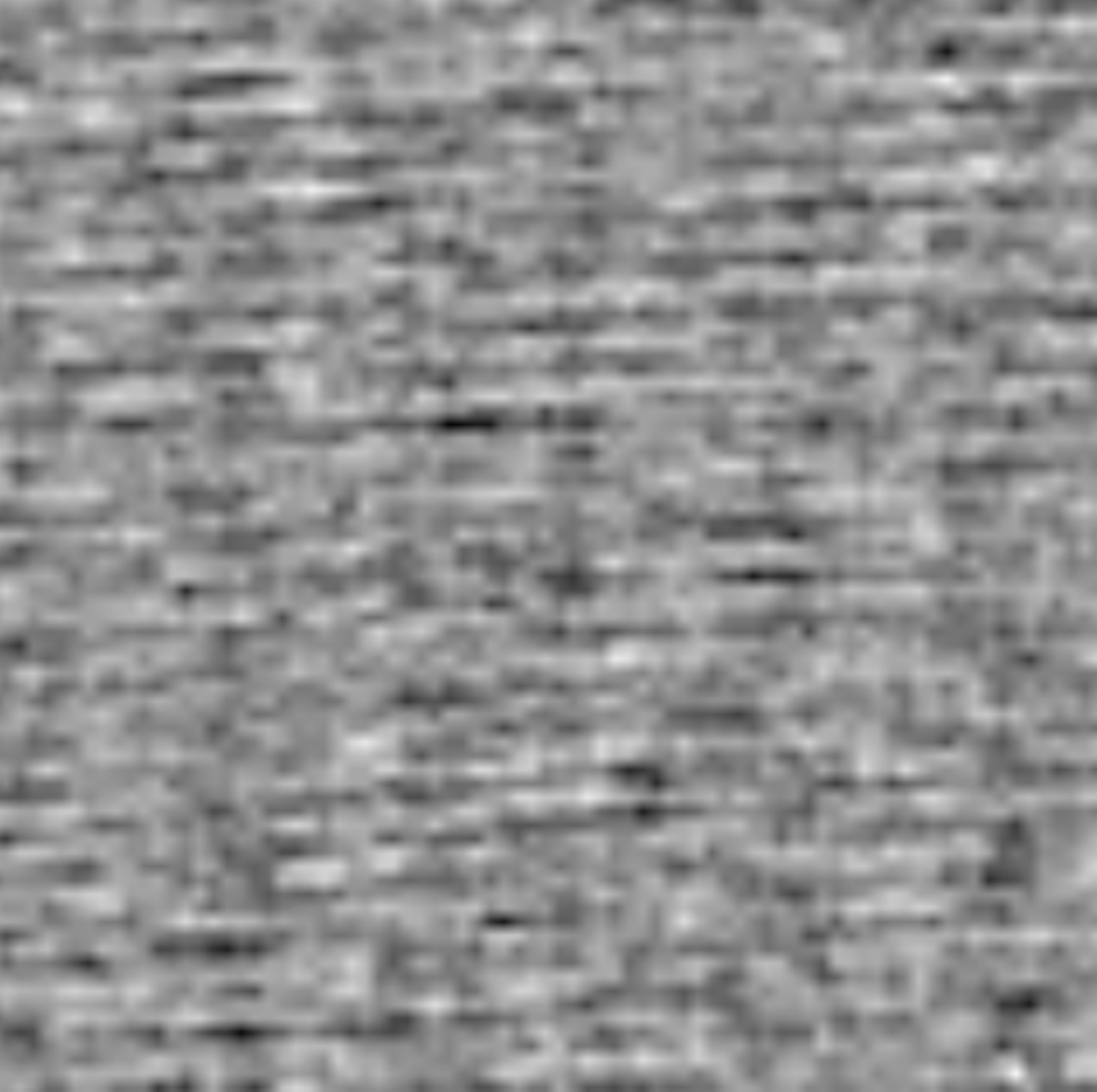}
\caption{Post-stack results for Mab}
\end{figure}

\begin{figure}[H]
\centering

\includegraphics[width=.9\linewidth]{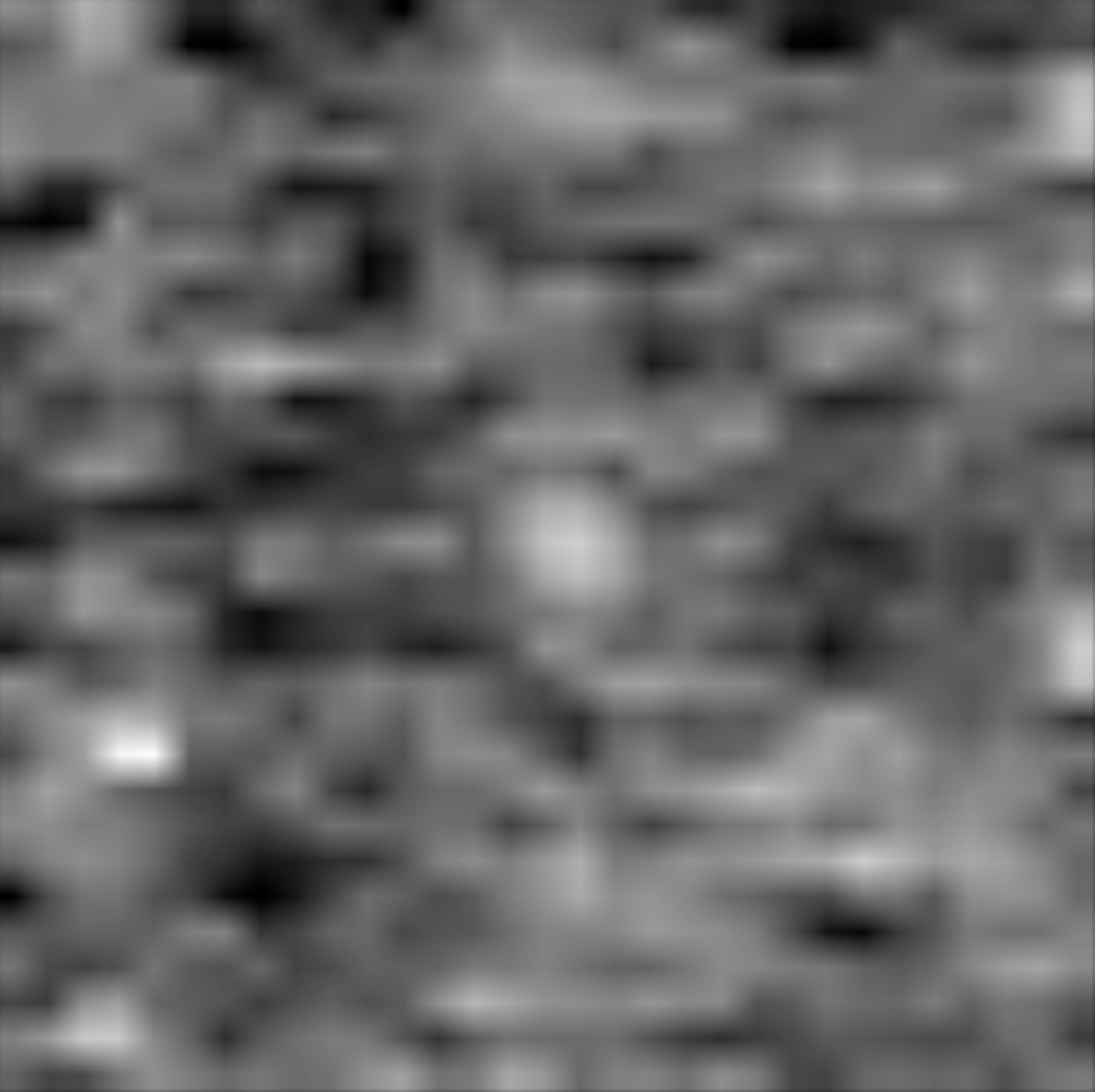}
\includegraphics[width=.9\linewidth]{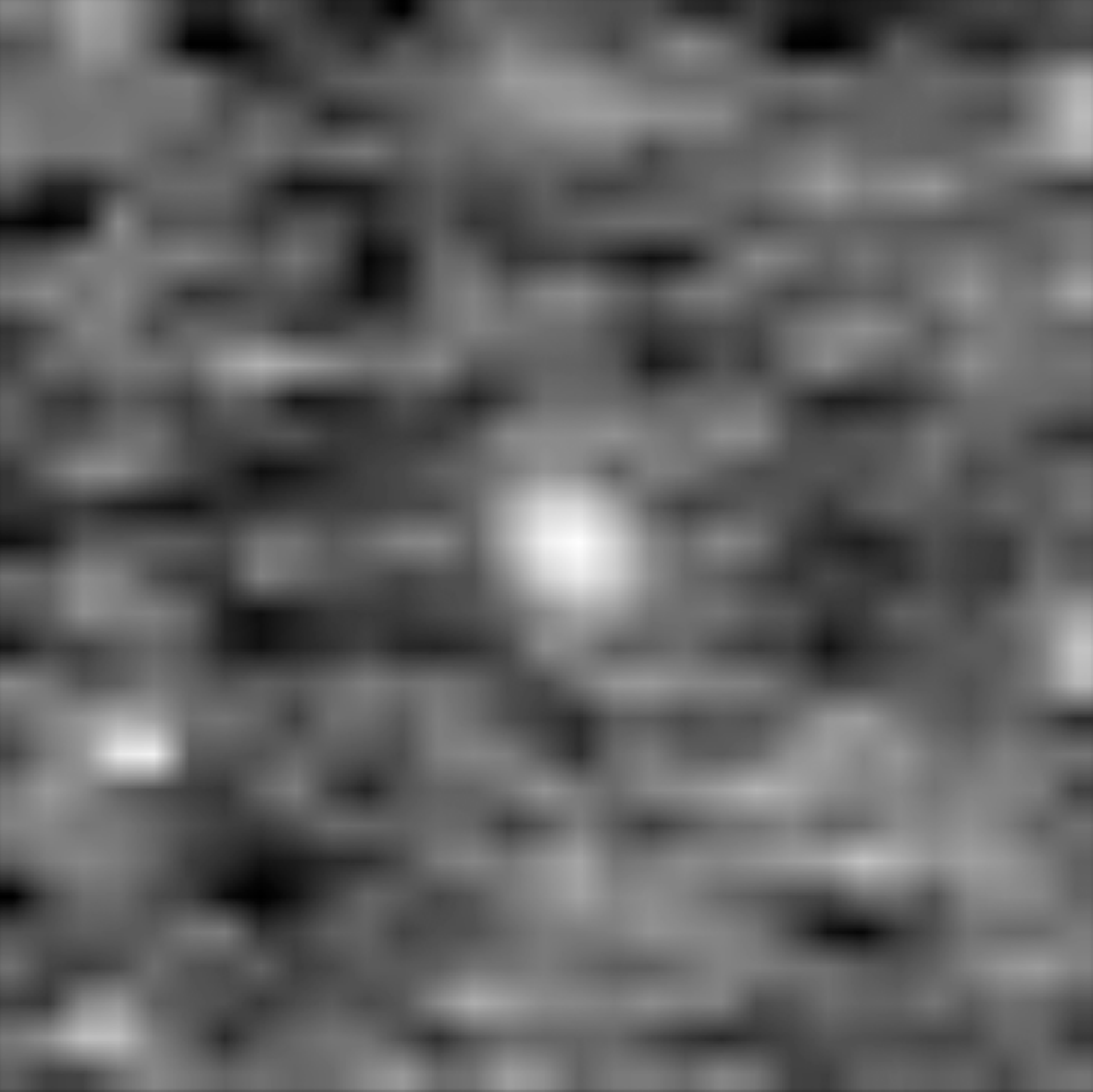}
\caption{Zoomed-in simulated 3$\sigma$ detection (top), 5$\sigma$ detection (bottom). The images are generated by matching the distribution of a point light source (calibrator star), but with the maximum value being the 3$\sigma$ or 5$\sigma$ detection value; taking an element wise maximum with this image and the original stacked image results in the images above.}
\end{figure}

Given we were unable to detect Mab, we can use our 3$\sigma$ upper limit to constrain Mab's radius or albedo, which are degenerate in this case. The albedo conveys information about the prevalence of absorbents at a specific wavelength, such as water ice or certain minerals.  For the given PSF as determined from our star observations, we can determine the 3$\sigma$ upper limit for an unresolved moon, which is $3.44 \times 10^{-16}$ erg s$^{-1}$ cm$^{-2}$/$\mu$m. With a solar flux of 185 erg s$^{-1}$ cm$^{-2}$/$\mu$m in \textit{H}-band at Uranus, we find a 3$\sigma$ upper limit to the \textit{I/F} of $\frac{1.86\times10^{-18}}{\Omega}$, with $\Omega$ the solid angle of Mab. Fig. 5 shows a simulated point source with an intensity equal to a 3σ or 5σ value. It acts as a simulation of Mab's appearance if it were to be detected in this dataset. 

Mab's orbit is in between those of Puck and Miranda. Puck, like the other small satellites, is relatively dark with an albedo of order 10\%, while Miranda is an icy moon, with a reflectivity equal to 46\% \citep{Showalter}. 

If Mab's radius is 12 km, as determined by \cite{Showalter} for an \textit{I/F} similar to that of Puck at visible wavelengths, our 3$\sigma$ upper limit would place an upper limit to the \textit{I/F} of 0.034 for Mab in the infrared, i.e., extremely dark. If Mab's radius is 6 km, as determined by \cite{Showalter} for an \textit{I/F} similar to that of Miranda at visible wavelengths, our 3$\sigma$ upper limit would place an upper limit to the \textit{I/F} of 0.136 for Mab in the infrared.

Hence, regardless of the precise \textit{I/F} in the visible, our data combined with the HST detection at visible wavelengths implies that Mab has a low \textit{I/F} at 1.6 $\mu$m compared to that at visible wavelengths, suggesting the surface is composed of materials that are absorbing in the IR, such as water-ice. We therefore suggest that Mab is an icy body, with an \textit{I/F} at visible wavelengths similar to that of Miranda (\textit{I/F} = 0.46; \cite{Showalter}), radius of 6 km, and an \textit{I/F} in the near-infrared of $\lesssim$ 0.14 (our 3$\sigma$ upper limit).
If Mab's \textit{I/F} at 2.2 $\mu$m would equal unity, \cite{dePater2006a} derived a 3$\sigma$ upper limit to Mab's radius of 3 km. For a 6 km radius object, the 3$\sigma$ upper limit to Mab's \textit{I/F} at 2.2 $\mu$m would be 0.25. A low \textit{I/F} at both 1.6 and 2.2 $\mu$m is highly suggestive of a surface covered with water ice, which has deep absorption bands at $\sim$1.5 and $\sim$2.0 $\mu$m. Indeed, large ratios in \textit{I/F} between visible and near-infrared may be most comparable to measurements of Pluto's small satellites Nix and Hydra, which show the presence of both amorphous and crystalline water ice \citep{Weaver, Cook}.

If Mab is indeed an icy moon, the blue $\mu$-ring, likely originating from micrometeorite bombardment on Mab, may be composed of icy grains much like the composition of Saturn's E-ring.

Finally, with Mab being more similar in composition to Miranda than to the inner moons, one might bring up the question of Mab's origin. Might it have been knocked off from Miranda during a collision, much like Hippocamp may have been a fragment of Proteus in the Neptunian system \citep{Showalter2}?

\quad

\quad

\section{Conclusion}
\label{S:5}
Photometric properties ofsixsatellites of Uranus are presented, based on 32 H-(1.4-1.8 $\mu$m) band images taken on August 29, 2015 from the Keck II Telescope on Maunakea, Hawaii with the near-infrared camera NIRC2 coupled to the adaptive optics system. We aligned the images using cross-correlation, and then determined the center of Uranus by minimizing the difference between the predicted location based on JPL ephemerides of the moons and the center of these moons as identified in the images. Using this center, we  mean-stacked the images of the moons to increase the signal-to-noise ratio. In comparing our derived Uranian moon reflectivities to previous measurements, we noticed that the small satellites in our data were significantly brighter than in previous observations. We investigated differences in both viewing geometries and phase angles, but neither can fully account for the discrepancy, and we conclude that there may be significant albedo variations between hemispheres. 

Despite our efforts to maximize the signal-to-noise ratio, we could not detect Mab. Instead, we attempted to identify which class of objects Mab belongs to by setting an upper limit on its albedo. \cite{Showalter} detected Mab in HST images, and derived a radius of 12 km if its albedo is similar to that of Puck, i.e., \textit{I/F} = 0.1. For such a moon, we derive an upper limit for the albedo of 0.034. If the \textit{I/F} of Mab would be similar to that of Miranda at visible wavelengths, i.e., 0.46, its radius would be 6 km. Our data then result in a 3$\sigma$ upper limit to the \textit{I/F} of 0.14. In either case the ratio in \textit{I/F} between the visible and near-infrared is about 3, which strongly suggests that Mab is an icy body like Miranda. We therefore suggest that Mab's radius is of order 6 km, and its albedo $\lesssim$ 0.45 in the visible, and $\lesssim$ 0.14 at 1.6 $\mu$m. 

Over the next years we will obtain an increasingly better view of the north pole of Uranus and its satellites. If our results are true, the reflectivities of the satellites will continue to increase, and hence we encourage continued observations of the Uranian satellite system. We also suggest to observe Mab with \textit{the James Webb Space Telescope} to obtain a reliable spectrum of this moon.

\section*{Acknowledgements}
This research was supported by NASA grant NNX16AK14G through the Solar System Observations (SSO) program to the University of California, Berkeley.

The data were obtained with the W.M. Keck Observatory, which is operated by the California Institute of Technology, the University of California, and the National Aeronautics and Space Administration. The Observatory was made possible by the generous financial support of the W.M. Keck Foundation.

Chris Moeckel was supported in part by the NRAO Student Observing Support (SOS) Program. Joshua Tollefson was supported by NASA Headquarters: under the NASA Earth and Space Science Fellowship program Grant NNX16AP12H to UC Berkeley. Samuel Paradis was participating through the Undergraduate Research Apprentice Program (URAP) at UC Berkeley.

The authors wish to recognize and acknowledge the very significant cultural role and reverence that the summit of Maunakea has always had within the indigenous Hawaiian community. We are most fortunate to conduct observations from this mountain.

\section*{ORCID iDs}
Samuel Paradis: \href{https://orcid.org/0000-0002-3581-312X}{0000-0002-3581-312X}

Chris Moeckel: \href{https://orcid.org/0000-0002-6293-1797}{0000-0002-6293-1797}

Josh Tollefson: \href{https://orcid.org/0000-0003-2344-634X}{0000-0003-2344-634X}

Imke de Pater: \href{https://orcid.org/0000-0002-4278-3168}{0000-0002-4278-3168
}



\end{document}